\documentclass{iau} 
\usepackage{graphicx}
\usepackage{hyperref}

\def\solphys{\textit{Solar~Phys.}}	
\def\ssr{\textit{Space~Sci.Rev.}}	

\title[Sunspot data collection of Specola Solare Ticinese] %% give here short title %%
{Sunspot data collection of Specola Solare Ticinese in Locarno}

\author[Renzo Ramelli et al.]   %% give here short author list %%
{Renzo Ramelli$^1$, 
Marco Cagnotti $^2$,  Sergio Cortesi$^2$, Michele Bianda$^1$, \and Andrea Manna$^2$}

\affiliation{$^1$Istituto Ricerche Solari Locarno \\ 
associated to Universit\`a della Svizzera italiana, \\
via Patocchi 57,
CH-6605, Locarno, Switzerland \\[\affilskip]
$^2$ Specola Solare Ticinese, \\
via ai Monti 146, CH-6605, Locarno, Switzerland }

\pubyear{2017}
\volume{340}  %% insert here IAU Symposium No.
\jname{Long-term datasets for the understanding of solar and stellar magnetic cycles}
\editors{D. Banerjee, 
J. Jiang, 
K. Kusano\&  
S. Solanki, eds.}
\begin{document}

\maketitle

\begin{abstract}

Sunspot observations and counting are carried out at the Specola
Solare Ticinese in Locarno since 1957 when it was built
as an external observing station of the Zurich observatory. 
When in 1980 the data center responsibility was transferred from ETH Zurich
to the Royal Observatory of Belgium in Brussels, the observations in Locarno continued and Specola Solare
Ticinese got the role of pilot station. The data collected at Specola cover now the last 6
solar cycles. 

The aim of this presentation is to discuss and give an overview
about the Specola data collection, the applied counting method and the future
archiving projects. The latter includes the publication of all data and
drawings in digital form in collaboration with the ETH Zurich University
Archives,
where a parallel digitization project is ongoing  
for the document of the former Swiss Federal Observatory in Zurich 
collected since the time of Rudolph Wolf.

\end{abstract}

\firstsection % if your document starts with a section,

\section{Historical overview}

The long tradition in Switzerland of
systematic observations and counting of sunspots was  started
by Rudolf Wolf in 1847 \cite[(Friedli, 2016)]{Friedli2016}, first in Bern and then in Zurich, where in 1855 he was 
nominated professor for astronomy both at the Zurich University and at the 
Swiss Federal Institute of Technology (ETH-Zurich). To study the solar cycle, 
that was discovered by Samuel Schwabe few years before, Wolf introduced the notorious
empirical index well know as {\it Wolf number} or as {\it Zurich relative
  sunspot number}, which is defined as:
\begin{equation}
\label{eqwolf}
R= k (10 g + s)
\end{equation}
where $g$ is the number of sunspot groups, $s$ the number of single
sunspots and $k$ a normalization factor which depends on the observer.
Being Wolf the reference, his observations were originally normalized with the
factor $k=1$.
The sunspot observations and the determination of the daily value of the Zurich relative
sunspot number continued at the Swiss Federal Observatory in Zurich 
with Wolf's successors Alfred Wolfer, William Brunner
and Max Waldmeier, who had a normalizing factor $k=0.6$ 
\cite[(Clette et al., 2014)]{2014SSRv..186...35C}. 

In 1936 the retired engineer Karl Rapp, grounder of BMW, moved to Locarno 
in Southern Switzerland and 
started regular observations of sunspots, collaborating with Brunner and
Waldmeier. The favorable weather conditions in Locarno allowed often to fill
the gaps of data when in Zurich clouds didn't allow to observe. Based on this positive
experience,  Waldmeier managed in the International Geophysical Year (1957) to ground
an external observing station in Locarno, that became the Specola Solare
Ticinese \cite[(Cortesi et al., 2016)]{Cortesi2016}. Sergio Cortesi and Araldo
Pittini were hired as Waldmeier's assistants, in order to carry out
observations at the Specola.

After Waldmeier's
retirement in 1979, ETH-Zurich decided to abandon the research activity related to the
sunspot counting. 
The Royal Observatory of Belgium in Brussels agreed to take
over this service. In January 1981, the new world data center in Brussels,
previously called SIDC and now
SILSO, took over the responsibility to determine the daily Wolf number, from
then on called {\it international sunspot number} (SSN) \cite[(Stenflo, 2016)]{Stenflo2016}. 
At that point,
thanks to experience gained with Waldmeier, Specola Solare Ticinese became the
pilot station with the role to guarantee the long term stability of the SSN. Observations
continued under the direction of Cortesi, that acted as main observer.
After some years of training, in 2011, Marco Cagnotti became the new main observer
at Specola.

%(Fig.\,\ref{fig1}).

%\begin{figure}[b]
%% \vspace*{-2.0 cm}
%\begin{center}
% \includegraphics[width=3.4in]{Path.eps} 
%% \vspace*{-1.0 cm}
% \caption{xxx}
%   \label{fig1}
%\end{center}
%\end{figure}

\section{The sunspot drawings of the Specola Solare Ticinese}

Observations in Locarno are obtained with a Coud\'e-Zeiss refractor
with an aperture of 15 cm. This is diaphragmed to 8 cm, in order to obtain a
better visual contrast.
The refractor projects a solar mirrored image with a diameter of 25 cm on a metallic
plate where a paper sheet is fixed, where the observer can draw the sunspots
with a pencil (Fig.~1). %\,\ref{fig1})
Originally, sunspot counting is done
following the rules taught by Waldmeier to the Specola observers
\cite[(Waldmeier, 1948)]{Waldmeier48}, where sunspots are weighted
according to their size. For instance, small point-like sunspots are counted as
1, larger sunspots as 2, a sunspot with penumbra of normal size is counted as
3, while a very large sunspot with penumbra is counted as 5.

\begin{figure}[htb]
% \vspace*{-2.0 cm}
\begin{center}
 \includegraphics[width=\linewidth]{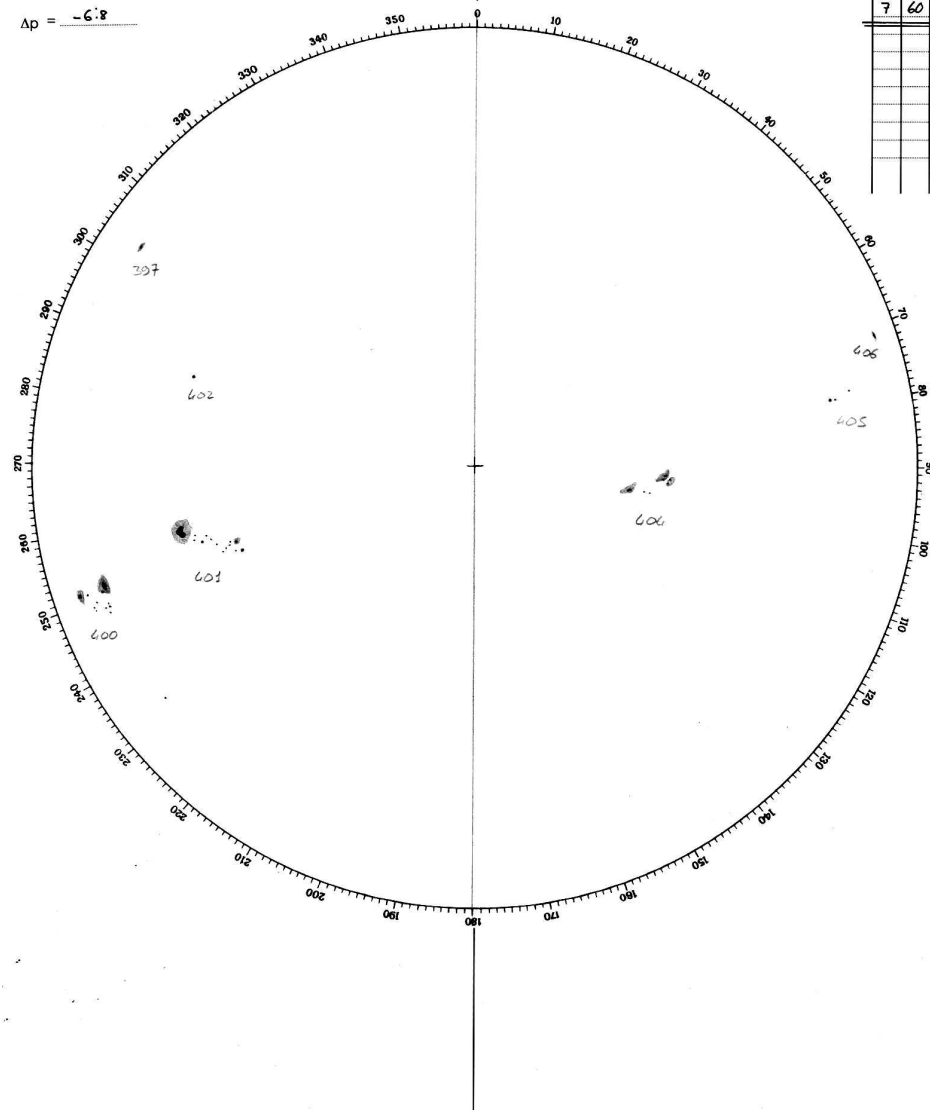} 
% \vspace*{-1.0 cm}
 \caption{Example of a sunspot drawing obtained at Specola Solare Ticinese in
   Locarno. The solar image is mirrored. In the top-right table, for each
   sunspot group, it
 is reported the serial number, the weighted counting of the sunspots, the
 Zurich classification and the unweighted counting.}
   \label{fig1}
\end{center}
\end{figure}

In 2012, it has been proposed to apply in parallel to the original counting
method, an unweighted counting where each sunspot is counted as 1
independently of its size. This allowed comparison studies of the two methods 
\cite[(Svalgaard et al., 2017)]{2017SoPh..292...34S} that were used in
recent works aiming at a better homogenization of the SSN in the past. 
In particular the results of this study led to a better understanding 
of the ``Waldmeier jump'' occurred around 1947 when
Waldmeier became the new director of the Swiss Federal Observatory in Zurich instead of Brunner.
These studies were taken into account
in the revision of the SSN introduced by SILSO on July 2015 by SILSO 
\cite[(Clette \& Lef{\`e}vre, 2016)]{2016SoPh..291.2629C}. One of the
decisions that was taken for this revision was to use the unweighted counting
of Locarno as the new reference\footnote{Another important change in the SSN
revision adopted by SILSO in July 2015 was to take Wolfer as reference, 
instead of Wolf for
the normalization factor $k=1$ present in equation \ref{eqwolf}.
\cite[(Clette \& Lef{\`e}vre, 2016)]{2016SoPh..291.2629C}}.  
Since August 2014 both weighted and unweighted countings of the Specola, are reported
officially on the sunspot drawings and communicated to SILSO.
Meanwhile, for further studies, we started to determine the unweighted
counting from the past drawings of the Specola. This work is still in progress.

\section{Sunspot data and drawings archiving}

The digitized Sunspot drawings of the Specola Solare Ticinese are published daily on the
WEB at the URL \url{http://specola.ch/e/drawings}. At the same address one
can find the archive that includes all digitized drawings made since 1981,
when the Specola became independent from ETH-Zurich. 
The drawings made at Specola from 1957 to 1980 are stored in the ETH Zurich
University Archives at the ETH-library, together with all the material
collected by the former Swiss Federal Observatory in Zurich. 

For the safe long term preservation of the sunspot drawings and data collected
in Locarno, Specola Solare Ticinese and  ETH Zurich University Archives are starting a project 
together. This foresees that all the drawings, including the ones made
previously than 1981, will be professionally digitized and made openly accessible
through the E-manuscripta digital platform for manuscript material from Swiss libraries and
archives accessible at \url{http://www.e-manuscripta.ch}. All drawings will have their own
digital object identifier (DOI). The digitization project of the Specola
drawings will complement a parallel project done by the ETH Zurich University Archives, in
which the full sunspot drawings collection of the former Swiss Federal
Observatory in Zurich is being digitized and made available on the same
digital platform E-manuscripta.

In addition, according to the project, it is foreseen 
that all the Specola drawings
made until now will be safely stored at the ETH Zurich University Archives.
Furthermore in the next years it is planned to prepare a digital sunspot group database that will be
published with open access. The database will contain information for each
sunspot group about observing date, observer, image quality, group ID serial number, Zurich
classification, latitude, original weighted counting. It is also foreseen to perform a
post-unweighted counting and to add this information to the database. 
Giving open access, we hope that the database will
be a useful data source for all researchers 
for further studies of the data series obtained at Specola.
%help the researchers to analyze the sunspot data series obtained at Specola.

\section{Acknowledgments}

The archiving and digitization project of the Specola data and drawings 
will be supported in the next years by the Federal Office of Meteorology and
Climatology MeteoSwiss, in the framework of GCOS Switzerland.
The regular research activity at Specola Solare Ticinese is mainly financed by
Canton Ticino through the Swisslos found. 

The authors acknowledge also the personnel of ETH Zurich University
Archives for the collaboration in the archiving and digitization 
 project, in particular Ms. Evelyn Boesch and Mr. Christian Huber.

\end{document}